\documentclass{llncs}
\pdfoutput=1

\usepackage{epsfig}
\usepackage{latexsym}
\usepackage{algorithmic}
\usepackage{algorithm}

\newcommand{\maxi}{\mbox{\footnotesize  max}}
\newcommand{\MST}{\mbox{\footnotesize  MST}}
\newcommand{\opt}{\mbox{\footnotesize  opt}}
\newcommand{\F}{\mathrm{I\kern-.23em F}}
\newcommand{\curr}{\mbox{\footnotesize curr}}
\newcommand{\currvert}{\mbox{\footnotesize pathv}}
\newcommand{\add}{\mbox{\footnotesize add}}
\newcommand{\NIL}{\mbox{NIL}}
\newcommand{\first}{\mbox{\footnotesize first}}
\newcommand{\deleteFirst}{\mbox{\footnotesize deleteFirst}}
\newcommand{\nex}{\mbox{\footnotesize next}}
\newcommand{\Path}{\mbox{\footnotesize Path}}

\newcommand{\pred}{\mbox{\footnotesize pred}}
\newcommand{\N}{\mathrm{I\kern-.23emI\kern-.29em N}}
\newcommand{\Q}{\mathrm{{\sf I}\kern-.48em Q}}
\newcommand{\Z}{\mathrm{Z\kern-.31em Z}}

  \title{Strategic deployment in graphs}
  
  \author{
       Elmar Langetepe$^1$
       \and Andreas Lenerz$^1$              
       \and Bernd Br\"uggemann$^2$  
}

\institute{University of Bonn, Germany,
         Institute of Computer Science I.        
\and 
        FKIE, Fraunhofer-Institute, Germany. 
        }

\date{\today}

\begin{document}
  \maketitle
  
  \begin{abstract}
  Conquerors of old (like, e.g., Alexander the Great or Ceasar) had to solve the following deployment problem. Sufficiently strong units had to be stationed at locations of strategic importance, and the moving forces had to be strong enough to advance to the next location.
To the best of our knowledge we are the first to consider the (off-line) graph version of this problem. While being NP-hard for general graphs, for trees the minimum number of agents and an optimal deployment can be computed in optimal polynomial time. Moreover, the optimal solution for the minimum spanning tree of an arbitrary graph $G$ results in a 2-approximation of the optimal solution for $G$.
  
  \end{abstract}

\section{Introduction}\label{intro-sec}
Let $G=(V,E)$ be a graph with non-negative edge end vertex weights $w_e$ and $w_v$, respectively.
We want to minimize the number of agents needed to traverse the graph subject to the following conditions. If vertex $v$ is visited for the first time, $w_v$ agents must be left at $v$ to cover it.
An edge $e$ can only be traversed by a force of at least $w_e$ agents. Finally, 
all vertices should be covered. All agents start in a predefined start vertex $v_s\in V$.
In general they can move in different groups. 
The problem is denoted as a \emph{strategic deployment problem} of $G=(V,E)$.

The above rules can also easily be interpreted for modern non-military applications.  
For a given network we would like to rescue or repair the sites (vertices) by 
a predifined number of agents, whereas traversing along the routes (edges) 
requires some escorting service. 
The results presented here can also be applied to a problem  
of positioning mobile robots for guarding a given terrain; see also~\cite{blls-mrgps-12}. 

 We deal with two variants regarding a \emph{notification} 
at the end of the task.  The variants are comparable to \emph{routes} (round-trips) and 
\emph{tours} (open paths) in traveling-salesman scenarios.

\begin{description}
\item[(Return)] Finally some agents have to return to the 
start vertex and report the success of the whole operation. 
\item[(No-return)] It suffices to fill the vertices as required, no agents 
have to return to the start vertex. 
\end{description}

Reporting the success in the return variant  means, 
that finally a set, $M$, of agents return to $v_s$ and the union of \emph{all} vertices 
visited by the members of $M$ equals $V$.

\begin{figure}[hbtp]%
  \begin{center}%
    \includegraphics[scale=0.7,keepaspectratio]{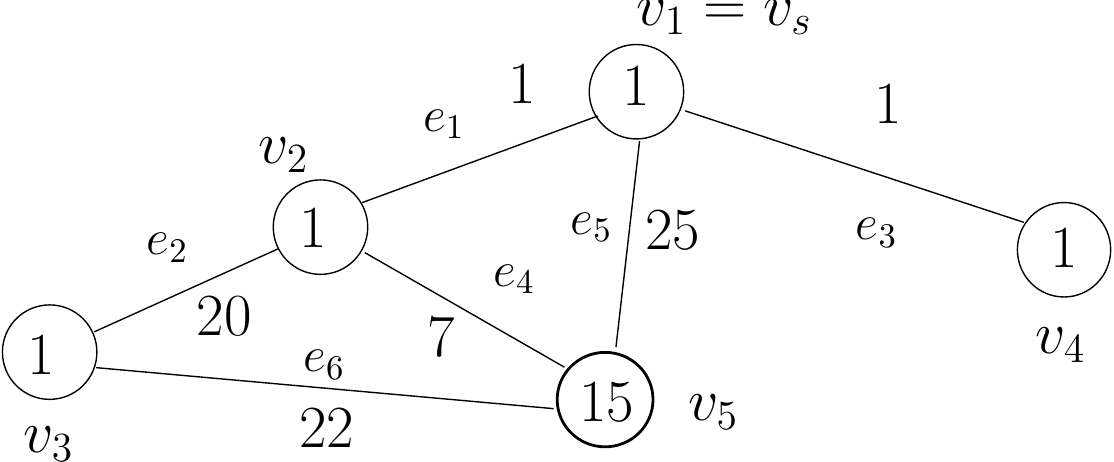}%
    \caption{A graph with edge and vertex weights. If the agents have to start at the vertex $v_1$ 
an optimal deployment strategy 
requires 23 agents and visits the vertices and edges in a single group in the order 
($v_1$, $e_1$, $v_2$, $e_2$, $v_3$, $e_2$, $v_2$, $e_1$, $v_1$, $e_3$, $v_4$, $e_3$,     $v_1$, $e_1$, $v_2$, $e_4$, $v_5$). 
The traversal fulfills the demand on the vertices in the 
order $v_1,v_2,v_3,v_4,v_5$ by the first visits w.r.t. the above sequence. At the end 
4 agents are not settled.}
    \label{example-fig}
  \end{center}%
\end{figure}
We give an example for the no-return variant for the graph of Figure~\ref{example-fig}. 
It is important that the first visit of a vertex immediately binds some units of the agents for the 
control of the vertex. For start vertex $v_s=v_1$ at least 23 agents are required. 
We let the agents run in a single group. In the beginning 
one of the agents has to be placed immediately in $v_1$. Then we traverse
edge $e_1$ of weight $1$ with 22 agents from $v_1$ to $v_2$. 
Again, we have to place one agent immediately at $v_2$. 
We move from $v_2$  to $v_3$ 
along  $e_2$ of weight 20 with 21 agents. After leaving one agent at $v_3$ 
we can still move back along edge $e_2$ (weight 20) 
from $v_3$  to $v_2$ with 20 agents.  The vertex $v_2$ was already covered before.
With 20 agents we now visit $v_4$  (by traversing $e_1$ (weight 1) and $e_3$ (weight 1), 
the vertex $v_1$ was already covered and can be passed without loss). 
We have to place one agent at $v_4$ and proceed
with 19 agents along $e_3$ (weight 1), $e_1$ (weight 1) and $e_4$ (weight 7) to $v_5$ 
where we finally have to place 15 agents. $4$ agents are not settled. \\
It can be shown that 
no other traversal requires less than 23 agents. 
By the results of Section~\ref{tree-sec} it turns out that the return variant solution has a different visiting order 
$v_1,v_2,v_3,v_5,v_4$ and requires 25 agents.  

Although the computation of an efficient flow of some items or 
 goods in a weighted network has a long tradition and has 
 been considered under many different aspects the 
 problem presented here cannot be covered 
 by known (multi-agent) routing, network-flow or agent-traversal problems.

  For example, in the classical \emph{transportation network} problem 
  there are source and sink nodes whose weights represent a 
  supply or a demand, respectively. The weight of an edge represents the 
  transportation cost along the edge. One would like to 
  find a transshipment schedule of minimum cost that fulfils 
  the demand of all sink nodes from the source nodes; see for example the monograph 
  of \cite{c-lp-83} and the textbooks \cite{kv-cota-07,amo-nftaa-93} . The solutions of such problems are often based 
  on linear programming methods for minimizing (linear) cost functions.  

  In a \emph{packet routing} scenario for a given weighted network, 
  $m$ packet sets each consisting of $s_i$ packets for $i=1,2,\ldots m$ are 
  located at $m$ given source nodes. For each packet set a specified 
  sink node is given. Here the edge weights represent an \emph{upper bound} on the 
  number of single packets that can be 
  transported along the edge in one time step. 
  One is for example interested in minimizing the so-called \emph{makespan}, i.e., the 
  time when the last packet arrives at its destination; see for example~\cite{psw-prca-09}.
  For a general overview see also the survey~\cite{ghks-prfcn-98}. 
  
  Similarily, in~\cite{lmkkktmj-amr-05} the \emph{multi-robot routing} problem considers 
   a set of agents that has to be moved from their start locations to the target locations. 
  For the movement between two locations a cost function is given and 
  the goal is to minimize the path costs.  Such multi-robot routing problems can be considered under many different constraints~\cite{sdi-ormam-10}. For the purpose of patrolling see the survey~\cite{pr-smrpa-11}.
  
  Additionally, \emph{online multi-agent traversal} problems in discrete environments have
  attracted some attention. The problem of exploring an \emph{unknown} 
  graph by a set of $k$ agents was considered for example in 
  \cite{dls-wrnm-07,fgkp-cte-06}.  
  Exploration means that at the end all vertices of the graph should have been visited.  
  In this motion planning setting either the goal is to optimize the number of 
  overall steps of the agents or to optimize the makespan, that 
  is to minimize the time when the last vertex is visited. 
 
  Some other work has been done for $k$~cooperative cleaners 
that move around in a grid-graph environment and have to \emph{clean} each 
vertex in a contaminated environment; see \cite{bkkl-asdsc-2011,wayb-ccsar-08}. 
In this model the task is different from 
a simple exploration since after a while \emph{contaminated} cells 
can reinfect cleaned cells. One is searching for strategies for 
a set of $k$~agents that guarantee successful cleanings.  

 Our result shows that finding the minimum number 
of agents required for the strategic deployment problem 
is NP-hard for general graphs even if all vertex weights are equal 
to one. In Section~\ref{NPhard-sec} this is shown by a 
reduction from 3-Exact-Cover (3XC). The optimal number of agents for 
the minimum spanning tree (MST) of the graph $G$ gives a $2$-approximation 
for the graph itself; see Section~\ref{tree-sec}. 
For weighted trees 
we can show that the optimal number of agents and a corresponding
strategy for $T$ can be computed in $\Theta(n \log n)$ time. Altogether, 
a $2$-approximation for $G$ can be computed efficiently. Additionally, 
some structural properties of the problem are given. 

The problem definition gives rise to many further interesting extensions. 
For example, here we first consider an offline version with global communication, but also 
online versions with limited communication might be of some interest. 
Recently,  we started to discuss the makespan or traversal time for a \emph{given} optimal number of agents. See for example the  masterthesis \cite{l-gn-12} 
supervised by the second author. 

\section{General graphs}\label{NPhard-sec}

We consider an edge- and vertex-weighted graph $G=(V,E)$. 
 Let $v_s\in V$ denote the start vertex for the 
traversal of the agents. 
W.l.o.g. we can assume that $G$ is connected and does not have multi-edges.

We allow that a traversal strategy subdivides the agents into groups 
that move separately for a while. A traversal strategy is a schedule for 
the agents. At any time step any agent decides 
to move along an outgoing edge of its current vertex towards  another vertex or the agent stays 
in its current vertex. We assume that any edge can be traversed
in one time step. \emph{Long} connections can be easily modelled by placing intermediate vertices of weight $0$ along the edge. 
Altogether, agent groups can arrive at some vertex $v$ at the same time from different edges.

The schedule is called \emph{valid} if the following condition hold. 
For the movements during a time step the number of agents that use 
a single edge has to exceed the edge weight $w_e$. After the movement 
for any vertex $v$ that already has been visited by some agents, the number of 
agents that are located at $v$ has to exceed the vertex weight $v_e$. 
From now on an \emph{optimal deployment strategy} is a valid schedule that 
uses the minimum number of agents required.

Let $N:=\sum_{v\in V} w_v$ denote
the number of agents required for the vertices in total. 
Obviously, the maximum overall edge weight $w_{\maxi}:=\max\{w_e|e\in E\}$ 
of the graph gives a simple upper bound for the additional agents (beyond $N$) 
used for edge traversals. 
This means that at most $w_{\maxi}+N$
agents will be required. With $w_{\maxi}+N$ agents one can 
for example use a DFS walk along the graph and let the agents run in a single group.

\subsection{NP-hardness for general graphs}\label{red3XC-sec}
\begin{figure}[hbtp]%
  \begin{center}%
    \includegraphics[scale=0.7,keepaspectratio]{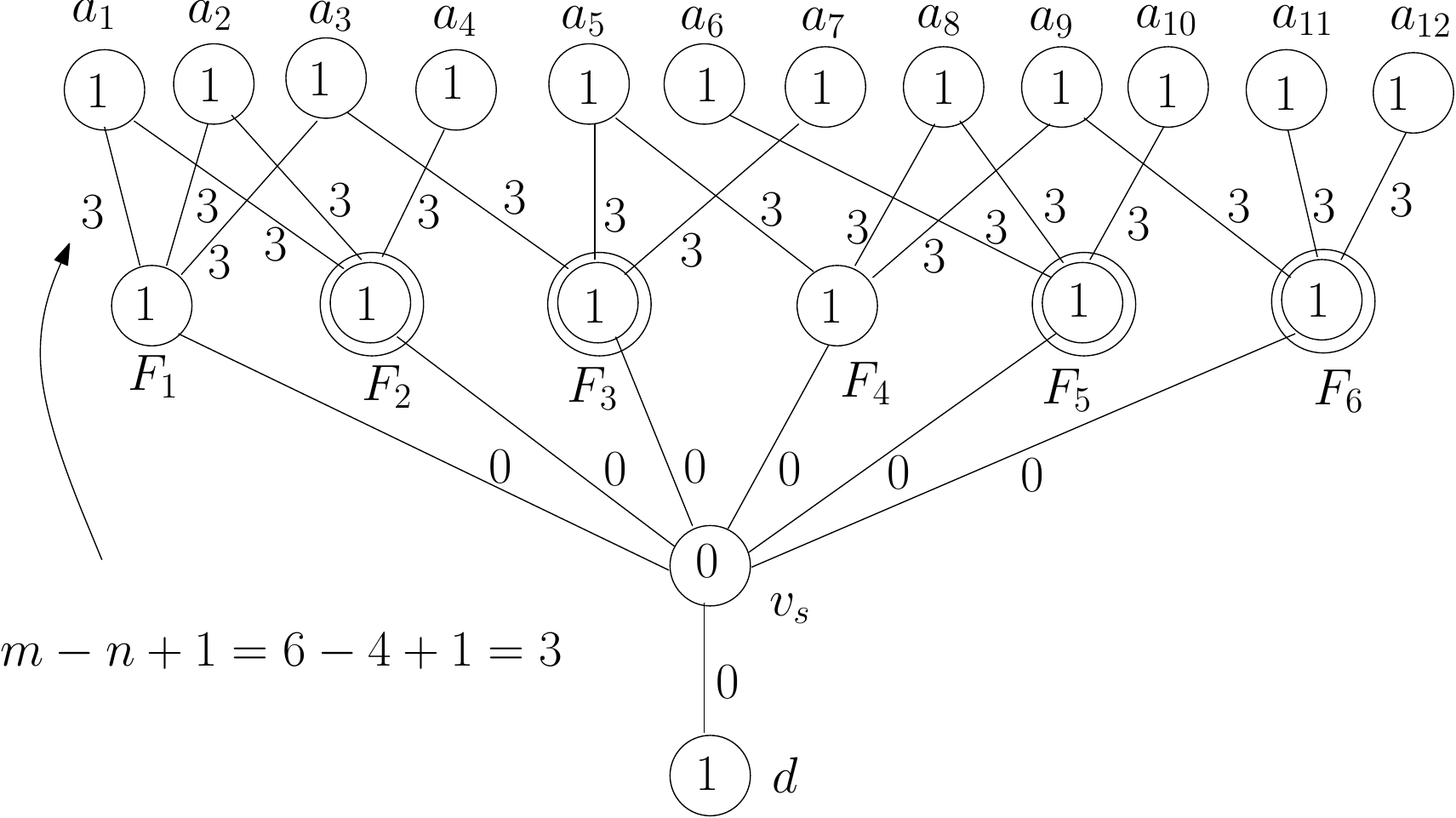}%
    \caption{For $X=\{a_1,a_2,\ldots,a_{12}\}$ and 
the subsets $\F=\{F_1,F_2,\ldots,F_6\}$ with $F_1=\{a_1,a_2,a_3\}$,
$F_2=\{a_1,a_2,a_4\}$, $F_3=\{a_3,a_5,a_7\}$,
$F_4=\{a_5,a_8,a_{9}\}$, $F_5=\{a_6,a_{8},a_{10}\}$ and 
 $F_6=\{a_9,a_{11},a_{12}\}$ there is an exact $3$-cover with 
$F_2$, $F_3$, $F_5$ and $F_6$. For the start vertex $v_s$ an optimal 
traversal strategy moves in a single group. We start with $3n + m + 1=19$ agents, first visit the vertices of $F_2$, $F_3$, $F_5$ and $F_6$ 
and cover all elements from there, visiting an element vertex last. 
After that $3n+n=4n=16$ agents have been placed and $m-n+1=3$ still have to be 
placed including the dummy node. With this number of agents we can move back along the corresponding 
edge of weight $m-n+1=3$ and place the remaining $3$ agents. 
}\label{Figures/ExampleRed-fig}
  \end{center}%
\end{figure}%
For showing that computing the optimal number of agents 
is NP-hard in general we make use of a reduction of 
the 3-Exact-Cover (3XC) problem. We give the proof for the no-return variant, first. 

The problem 3-Exact-Cover (3XC) is given as follows. 
Given a finite ground set $X$ of $3n$ items and a set $\F$ of 
subsets of $X$ so that any $F\in \F$ contains exactly 
$3$ elements of $X$. 
The decision problem of 3XC is defined as follows: 
Does $\F$ contain an exact cover of $X$ of size $n$? More precisely 
is there a subset $F_c\subseteq \F$ so that the collection 
$F_c$ contains all elements of $X$ and $F_c$ consists of precisely $n$ 
subsets, i.e. $|F_c|=n$. It was shown by Karp that
this problem is NP-hard; see Garey and Johnsson\cite{gj-cigtn-79}.

Let us assume that such a problem is given. We  define the 
following deployment problem for $(X,\F)$. 
Let $X=\{a_1,a_2,\ldots,a_{3n}\}$. For any 
$a_i$ there is an \emph{element vertex} $v(X)_i$ of weight $1$. 
Let $\F$ consists of $m\geq n$ subsets of size $3$, say
$\F=\{F_1,F_2,\ldots,F_m\}$.
For any $F_j=\{a_{j_1},a_{j_2},a_{j_3}\}$ we define a 
\emph{set vertex} $v(\F)_j$ of weight $1$ and we insert three
edges $(v(\F)_j,v(X)_{j_1})$, $(v(\F)_j,v(X)_{j_2})$ and 
$(v(\F)_j,v(X)_{j_3})$ each of weight $m-n+1$. 
Additionally, we use a sink vertex $v_s$ of weight $w_{v_s}=0$ 
and insert $m$ edges $(v_s,v(F)_j)$ from the sink 
to the set vertices of $\F$. All these edges get weight $0$. 
Additionally, one 
dummy node $d$ of weight $w_d=1$ is added as well as an 
edge $(v_s,d)$ of weight $0$. 
 
Figure \ref{Figures/ExampleRed-fig} shows an example of the 
construction for the set $X=\{a_1,a_2,\ldots,a_{12}\}$ and 
the subsets $\F=\{F_1,F_2,\ldots,F_6\}$ with $F_1=\{a_1,a_2,a_3\}$,
$F_2=\{a_1,a_2,a_4\}$, $F_3=\{a_3,a_5,a_7\}$,
$F_4=\{a_5,a_8,a_{9}\}$, $F_5=\{a_6,a_{8},a_{10}\}$ and 
 $F_6=\{a_9,a_{11},a_{12}\}$ with $m-n+1=3$. 

Starting from the sink node $v_s$ we are asking whether there is 
an agent traversal schedule that requires exactly $N=3n + m +1$ agents. 
If there is such a traversal this is optimal (we have to fill all vertices). 
The following result holds. 
\emph{If and only if $(X,\F)$ has an exact 
$3$-cover, the given strategic deployment problem can be solved with 
exactly $N=3n + m +1$ agents.} 

Let us first assume that an exact $3$-cover exists. In this case we 
start with $N=3n + m +1$ agents at $v_s$ and let the agents run in a single 
group. 
First we successively visit the set vertices that build the cover 
and fill all $3n$ element vertices using $3n + n$ agents in total. 
More precisely, for the set vertices that build the cover 
we successively enter such a vertex from $v_s$, place one agent there and 
fill all three element vertices by moving back and forth along the corresponding edges. 
Then we move back to $v_s$ and so on.  At any such operation the set of agents 
is reduced by $4$. Finally, when the last set vertex of the cover was visited, 
we end in the overall last element vertex. After fulfilling the 
demand there, we still have 
$N-4n=3n + m + 1 - 4n = m-n + 1$ agents for traveling back to $v_s$ along the corresponding 
edges.  Now we fill the remaining set vertices by successively moving forth and back 
from $v_s$ along the edges of weight $0$. Finally, with the last agent, we can visit and fill 
the dummy node.

Conversely, let us assume that there is no exact $3$-cover for $(X,\F)$ 
and we would like to solve the strategic deployment problem with 
$N=3n + m +1$ agents. At some point an optimal solution for the strategic 
deployment problem has to visit the last element vertex $v(X)_j$, starting 
from a set vertex $v(\F)_i$. Let us assume that we are in $v(\F)_i$ and 
would like to move to  $v(X)_j$ now and $v(X)_j$ was not visited before. 
Since there was no exact $3$-cover we
have already visited strictly more than $n$ set vertices at this point and 
exactly $3n-1$ element vertices have been visited. This  means at 
least $3n-1+n+1=4n$ agents have been used. 

Now we consider two cases. If the dummy node was already visited, 
starting with $N$ agents we only have 
at most $3n + m + 1 - 4n - 1 = m - n$ agents for travelling toward 
the last element vertex, this means that we require an 
additional agent beyond $N$ for traversing the edge of weight $m - n +1$. 
If the dummy node was not visited before and we now decide to move to the last element vertex, 
we have to place one agent there. 
This means for travelling back from the last element vertex along some edge (at least the dummy must 
still be visited), we still require $m-n+1$ agents. Starting with $N$ at the beginning 
at this stage only $3n + m + 1 - 4n - 1 = m - n$ are given. At least one 
additional agent beyond $N$ is necessary for travelling back to the dummy node 
for filling this node. 

Altogether, we can answer the $3$-Exact-Cover decision problem 
by a polynomial reduction into a strategic deployment problem. 
The proof also works for the return variant, where at least one agent has to 
return to $v_s$, if we omit the dummy node, make use of $N:=3n+m$ and set the non-zero weights to 
$m-n$.

\begin{theorem}\label{NPhard-theo} 
Computing the optimal number of agents for the strategic deployment problem of a 
general graph $G$ is NP-hard. 
\end{theorem} 

\subsection{2-approximation by the MST}\label{2approx-sec}

For a general graph $G=(V,E)$ we consider its minimum spanning tree 
(MST) and consider an optimal deployment strategy on the MST. 

\begin{lemma}\label{MSTApprox-theo} 
An optimal deployment strategy for the minimum spanning tree (MST) of 
a weighted graph $G=(V,E)$ gives a $2$-approximation of the optimal 
deployment strategy of $G$ itself.
\end{lemma} 

\paragraph{Proof:} 
Let $e$ be an edge of the MST of $G$ with maximal weight $w_e$ 
among all edges of the MST.
It is simply the nature of the MST, that any traversal 
of the graph that visits all vertices, has to use an edge of weight at least $w_e$.  
The optimal deployment strategy has to traverse an 
egde of weight at least $w_{e}$ and 
requires at least $k_{\opt}\geq \max\{N,w_{e}\}$ 
agents. The optimal strategy for the MST approach requires 
at most $k_{\MST}\leq w_{e}+N$ agents which  
results to the bound $k_{\MST}\leq 2 k_{\opt}$.~\qed

\subsection{Moving in a single group}\label{onegroup-sec}
In our model it is allowed that the agents run in different groups. 
For the computation of the optimal number of agents required, this is not necessary. 
Note that \emph{group-splitting} strategies are necessary for minimizing the \emph{completion time}. 
Recently,  we also started to discuss such optimization criteria;  
see the  masterthesis \cite{l-gn-12}  supervised by the first author. 

During the execution of the traversal there is a set 
of \emph{settled} agents  that already had to be placed
at the visited vertices and a set of \emph{non-settled} agents 
that still move around.  We can show that the non-settled agents can 
always move in a single group. For simplicity we give a proof for trees.

\begin{theorem}\label{groupTree-thm} 
For a given weighted  tree $T$ and the given minimal number of agents required,  
there is always a deployment strategy that lets all non-settled agents move in a single group. 
\end{theorem}

 \paragraph{Proof:} We can reorganize any optimal strategy accordingly, so
 that the same number of agents is sufficient. 

 Let us assume that at a vertex $v$ a set of agents $X$ is separated into two 
 groups $X_1$ and $X_2$ and they separately explore disjoint parts $T_1$ 
 and $T_2$ of the tree. Let $w_{T_i}$ be the maximum edge weight of the 
 edges traversed by the agents $X_i$ in $T_i$, respectively.
 Clearly $|X_i|\geq w_{T_i}$ holds.
Let $|X_1|\geq |X_2|$ hold and let $X_2'$ be the set of non-settled agents 
 of $X_2$ after the exploration of $T_2$. 
 We can explore $T_2$ by $X=X_1\cup X_2$ agents first, and we do not need the set 
 $X_2$ there. $|X_1|\geq w_{T_2}$ means that we can move back with   
 $X_1\cup X_2'$ agents to $v$ and start the exploration for $T_1$.

 The argument can be applied successively for any split of a group. 
  This also means that we can always collect all non-settled vertices in 
  a single moving group.~\qed
 
Note that the above Theorem also holds for general graphs $G$. The general proof 
requires some technical details because a single vertex might collect agents from 
different sources at the first visit. We omit the rather technical proof here.

\begin{proposition}\label{group-thm} 
For a given weighted  graph $G$ and the given minimal number of agents required,  
there is always a deployment strategy that lets all non-settled agents move in a single group. 
\end{proposition}

\subsection{Counting the number of agents}\label{SimpleAlg-sec}

From now on 
we only consider strategies where the non-settled agents always move in a single group.
Before we proceed, we briefly explain how the number of agents 
can be computed for a strategy  
given by a sequence $S$ of vertices and edges that 
are visited and crossed successively. A pseudocode is given in Algorithm~\ref{eval-alg}.
The simple counting procedure will be adapted  for Algorithm~\ref{eval2-alg} in Section~\ref{Return-sec} for 
counting the optimal number of agents efficiently.
\begin{algorithm}
\caption{Number of agents, for $G=(V,E)$ and given sequence $S$ 
of vertices and edges.}
\label{eval-alg}
\begin{algorithmic}
\STATE $N:=\sum_{v\in V} w_v$; $\curr:=N$; $\add:=0$; $x:=\first(S)$;

\WHILE {$x \neq \NIL$}
\IF {$x$ is an edge $e$} 
        \IF {$\curr< w_e$} \STATE $\add:=\add+(w_e-\curr)$; $\curr:=w_e$; \ENDIF
\ELSIF {$x$ is a vertex $v$} 
        \IF {$\curr < w_v$} \STATE $\add:=\add+(w_v-\curr)$; $\curr:=0$; 
        \ELSE \STATE $\curr:=\curr-w_v$;
        \ENDIF
        \STATE $w_v:=0$;
\ENDIF
\STATE $x:=\nex(S)$;
\ENDWHILE
\STATE \mbox{\bf\small RETURN} $N+\add$
\end{algorithmic}
\end{algorithm}

For a sequence $S$ of vertices and edges that are visited and crossed 
by  a single group of agents the required number of agents can be computed 
as follows. We count the number of additional agents 
beyond $N$ (where $N$ is the overall sum of the vertex weights) in a variable 
$\add$. In another variable $\curr$ we count the number of agents currently available. 
In the beginning $\add:=0$ and $\curr:=N$ holds. A strategy successively 
crosses edges and visits vertices of the tree, this is given in the sequence $S$. 
We always choose the next element $x$ (edge $e$ or vertex $v$) out of the sequence. 
If we would like to cross an edge $e$, we check whether $\curr\geq w_e$ holds. If not we set $\add:=\add+(w_e-\curr)$ and 
$\curr:=w_e$ and can cross the edge now. If we visit a vertex $v$ we similarily 
check whether $\curr\geq w_v$ holds. If this is true, we set $\curr:=\curr-w_v$. 
If this is not true, we set $\add:=\add+(w_v-\curr)$ and $\curr:=0$. In any case we set $w_v:=0$, 
the vertex is filled after the first visit. Obviously this simple 
algorithm counts the number of agents required in the number of traversal steps 
of the single group.

\section{Optimal solutions for trees}\label{tree-sec}
Lemma \ref{MSTApprox-theo}  suggests that for 
a 2-approximation for a graph $G$, 
we can consider its MST. Thus, it makes sense to solve the problem 
efficiently for trees. Additionally, by Theorem \ref{group-thm} it suffices 
to consider strategies of single groups.

\begin{figure}[hbtp]%
  \begin{center}%
    \includegraphics[scale=0.5,keepaspectratio]{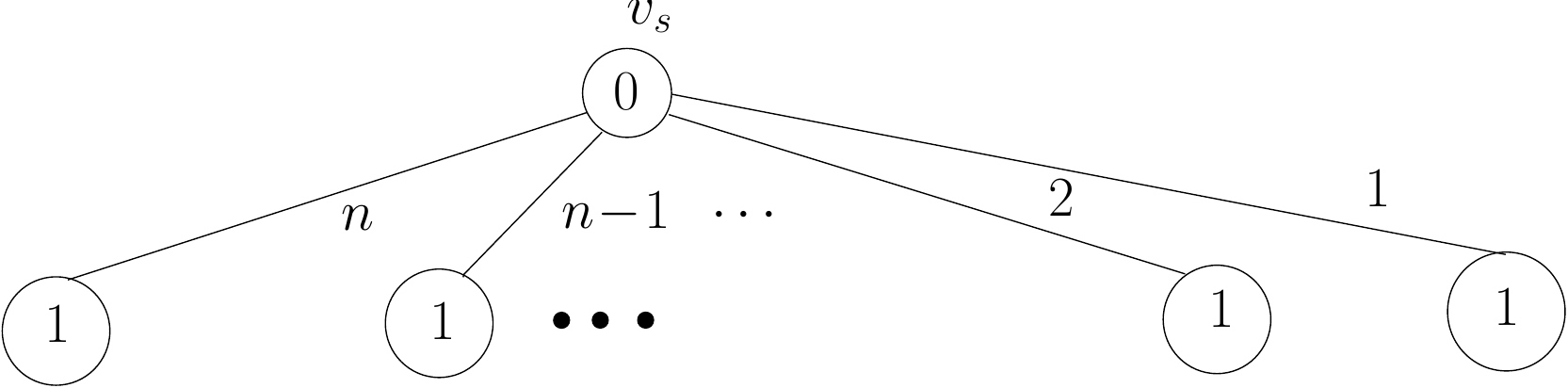}%
    \caption{
An optimal strategy that starts and ends in $v_s$ has to visit the leafs 
with respect to the decreasing order of the edge weights. The minimal number of  agents is $n+1$. Any other order will lead to at least one extra agent.}
    \label{lowerboundN-fig}
  \end{center}%
\end{figure}

\subsection{Computational lower bound}\label{lbtree-sec}
Let us first consider the tree in Figure \ref{lowerboundN-fig} and the return variant. 
Obviously it is possible to use $n+1$ agents and visit the edges in the 
decreasing order of the edge weights $n,n-1,\ldots,1$. Any other order will increase
the number of agents. If for example in the first step an edge of weight $k\neq n$ 
is visited, we have to leave one agent at the corresponding vertex. Since the 
edge of weight $n$ still has to be visited and we have to return to the start, 
$n+1$ agents in total will not be sufficient. So first the edge of weight $n$ has to 
be visited. This argument can be applied successively.

Altogether, by the above example there seems to be a computational 
lower bound for trees with respect to sorting the edges by their weights. 
Since integer values can be sorted by bucket sort in linear time, such a lower bound 
can only be given  for real edge and vertex weights. This seems to be a natural 
extension of our problem. We consider the transportation of sufficient material 
along an edge (condition 1.). Additionally, the demand of a 
vertex has to be fully satisfied before transportation can go on (condition 2.). 
How many material is required?

For a computational lower bound for trees we consider 
 the Uniform-Gap problem. Let us assume that $n$ unsorted real numbers  $x_1,x_2,\ldots,x_n$ and an $\epsilon>0$ are given. 
Is there a permutation $\pi:\{1,\ldots,n\}\rightarrow\{1,\ldots,n\}$ so that 
$x_{\pi(i-1)}=x_{\pi(i)}+\epsilon$ for $i=2,\ldots,n$ holds? 
In the algebraic decision tree model this problem has computational time bound $\Omega(n \log n)$;
see for example~\cite{ps-cgi-85}. 

In Figure \ref{lowerboundN-fig} we simply replace the vertex weights of $1$ by $\epsilon$ 
and the $n$ edge weights by $x_1,x_2,\ldots,x_n$. 
With the same arguments as before we conclude: 
If and only if the Uniform-Gap property holds, a unique optimal strategy has to
visit the edges in a single group in the order of decreasing edge weights $x_{\pi(1)}>x_{\pi(2)}>\cdots>x_{\pi(n)}$ 
and requires an amount of $x_{\pi(1)}+\epsilon$ in total. 
Any other order will lead to at least one extra $\epsilon$. 

The same arguments can be applied to the no return variant by simple modifications. Only the vertex weight of the smallest $x_j$, say $x_{\pi(n)}$,
is set to $x_{\pi(n)}$. 

\begin{lemma}\label{lower-lem} 
Computing an optimal deployment strategy for a tree of size $n$ with positive real 
edge and vertex weights takes $\Omega(n \log n)$ computational time 
in the algebraic decision tree model. 
\end{lemma}

\subsection{Collected subtrees}\label{Algorithm-sec}
%
The proof of Lemma~\ref{lower-lem} suggests  
to visit the edges of the tree in the order of decreasing weights. 
For generalization we introduce the following notations for a tree $T$ with root vertex $v_s$.

For every leaf $b_l$ along the unique shortest path, 
$\Pi_{v_s}^{b_l}$, from the root $v_s$ to $b_l$ there is an edge $e(b_l)$ with 
weight $w_{e(b_l)}$, 
so that $w_{e(b_l)}$ is greater than or equal to any other edge weight along 
$\Pi_{v_s}^{b_l}$. Furthermore,  we choose $e(b_l)$ so that it 
has the shortest edge-distance 
to the root among all edges with the same weight. Let $v(b_l)$ denote 
the vertex of $e(b_l)$ that is closer to the leaf $b_l$. 
Thus, every leaf $b_l$ defines a unique path, $T_{b_l}$, from $v(b_l)$ to 
the leaf $b_l$ with incoming edge $e(b_l)$ with edge weight $w_{e(b_l)}$. 
The edge $e(b_l)$ \emph{dominates} the leaf $b_l$ and also the  
path $T(b_l)$.

\begin{figure}[hbtp]%
  \begin{center}%
    \includegraphics[scale=0.53,keepaspectratio]{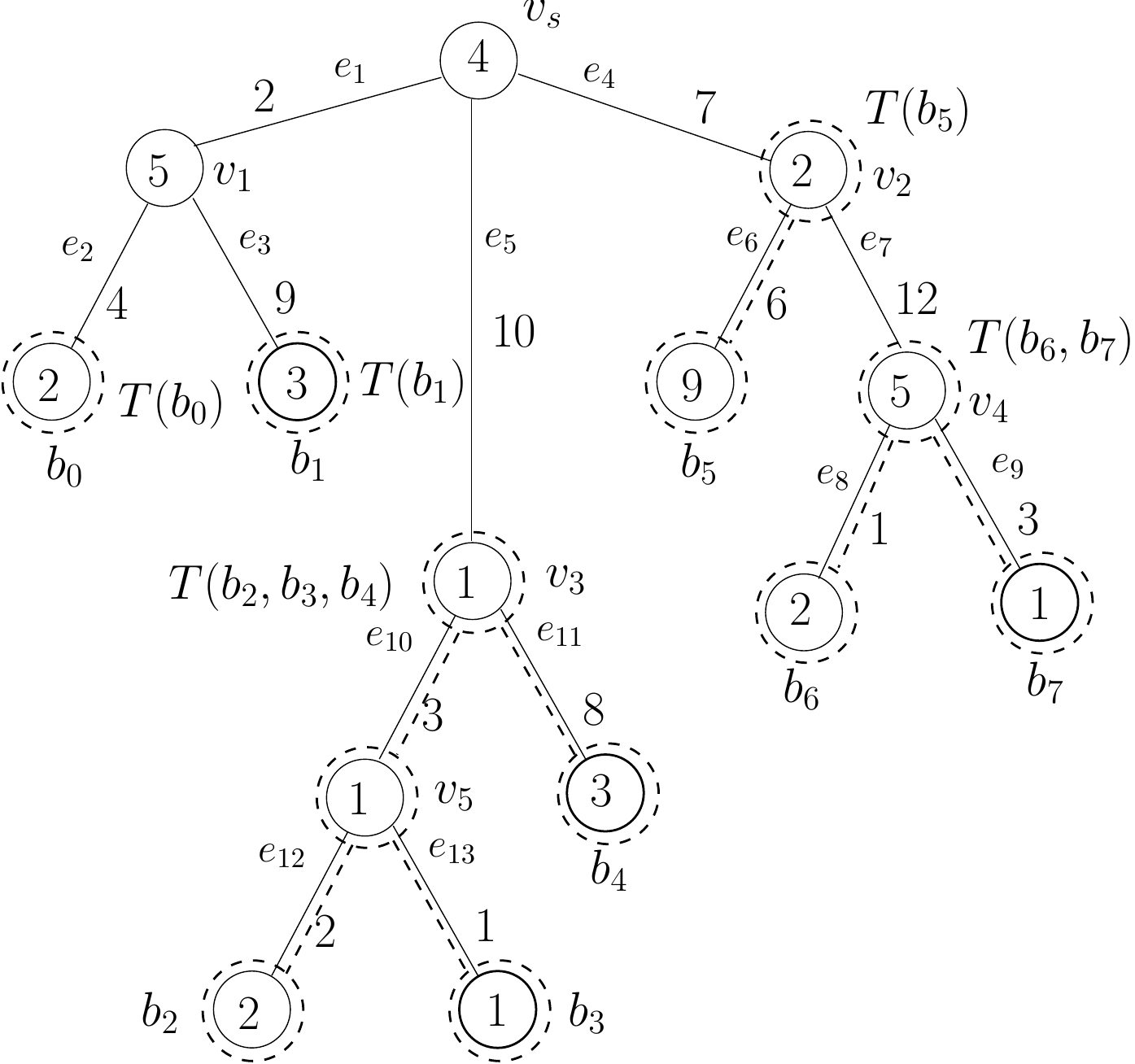}%
    \caption{The optimal strategy with start and end vertex $v_s$ visits, 
fully explores and leaves the 
collected subtrees $T(b_6,b_7)$, $T(b_2,b_3,b_4)$, $T(b_1)$, $T(b_5)$ and  
$T(b_0)$ in the order of the weights $w_{e_7}=12$, $w_{e_5}=10$, $w_{e_3}=9$,
$w_{e_4}=7$ and $w_{e_2}=4$ of the dominating edges.}
    \label{exampleII-fig}
  \end{center}%
\end{figure}

For example in Figure~\ref{exampleII-fig} we have $e(b_2)=e_5$ and 
$v(b_2)=v_3$, the path $T(b_2)$ from $v_3$ over $v_5$ to $b_2$ 
is dominated by the edge $e_5$ of weight $10$. 

If some paths $T_{b_{l_1}},T_{b_{l_2}},\ldots, T_{b_{l_m}}$ are dominated 
by the same edge $e$, we collect all those paths in a \emph{collected subtree} denoted by
$T(b_{l_1},b_{l_2},\ldots,b_{l_m})$. The tree has  
unique root $v(b_{l_1})$ and is dominated by unique edge $e(b_{l_1})$.

For example, in Figure~\ref{exampleII-fig} for $b_6$ and $b_7$
we have $v(b_6)=v(b_7)=v_4$ and $e(b_6)=e(b_7)=e_7$ and 
$T(b_6,b_7)$ is given by the tree $T_{v_4}$ that is 
dominated by edge $e_7$. 

Altogether, for any tree $T$ there is a unique set of 
disjoint collected subtrees (a path is a subtree as well) as uniquely defined above 
and we can sort them by the weight of its dominating edge. 
For the tree in  Figure~\ref{exampleII-fig} we have disjoint subtrees
$T(b_6,b_7)$, $T(b_2,b_3,b_4)$, $T(b_1)$, $T(b_5)$ and  
$T(b_0)$ in this order.

\subsection{Return variant for trees}\label{Return-sec}
%
%
We show that the collected subtrees can be visited in the order of the dominating edges.

\begin{theorem}\label{TreeBack-theo} 
An optimal deployment strategy that has to start and end at the same root vertex $v_s$ 
of a tree $T$ can visit the disjoint subtrees $T(b_{l_1},b_{l_2},\ldots, b_{l_m})$ 
in the decreasing order of the dominating edges. 

Any tree $T(b_{l_1},b_{l_2},\ldots, b_{l_m})$ 
can be visited, fully explored in some order (for example by DFS) 
and left then.
 
An optimal visiting order of the leafs and the optimal number of 
agents required can be computed in $\Theta(n \log n)$ time for real 
edge and vertex weights and in optimal $\Theta(n)$ time for integer weights.
\end{theorem} 

For the proof of the above Theorem we first show that we can reorganize any 
optimal strategy so that at first the tree $T(b_{l_1},b_{l_2},\ldots, b_{l_m})$ 
with maximal dominating edge weight can be visited, fully explored and 
left, if the strategy does not end in this subtree (which is always true for 
the return variant).  
The number of 
agents required cannot increase. This argument can be applied successively.
Therefore we formulate the statement in a more general fashion. 

\begin{lemma}\label{orderbyedge-lem} 
Let $T(b_{l_1},b_{l_2},\ldots, b_{l_m})$  be a subtree that is dominated 
by an edge $e$ which has the greatest weight among all edges that dominate a subtree. 

Let $S$ be an optimal deployment strategy that visits some vertex $v_t$ last 
and let  $v_t$ be not a vertex inside the tree $T(b_{l_1},b_{l_2},\ldots, T_{b_{l_m}})$.
The strategy $S$ can be reorganized so that first the 
tree $T(b_{l_1},b_{l_2},\ldots, b_{l_m})$ can be visited, fully explored 
in any order and finally left then.
\end{lemma}

\paragraph{Proof:} The tree $T(b_{l_1},b_{l_2},\ldots, T_{b_{l_m}})$ rooted at $v(b_{l_1})$ and 
with maximal dominating edge weight $w_{e(b_{l_1})}$ 
does not contain another subtree $T(b_{k_1},b_{k_2},\ldots, T_{b_{k_n}})$. 
This means that $T(b_{l_1},b_{l_2},\ldots, T_{b_{l_m}})$ is the full 
subtree $T_{v(b_{l_1})}$ of $T$ rooted at $v(b_{l_1})$.
Let $\Path(v(b_{l_1})$ denote the number of agents that has to be settled along the unique path from $v_s$ to the predecessor, $\pred(v(b_{l_1}))$, of $v(b_{l_1})$.

Let us assume that an optimal strategy is given by a sequence $S$ 
and let $S_v(i)$ denote the strategy that ends after the $i$-th visit of 
some vertex  $v$ in the sequence of $S$. Let $|S_v(i)|$ denote
the number of settled agents and let $\curr(S_v(i))$ denote the number of non-settled agents after the $i$-th visit of 
$v$. We would like to replace $S$ by a sequence $S'S''$. 
If vertex $v(b_{l_1})$ is finally visited, say for the $k$-th time, in the sequence $S$, we require $\curr(S_{v(b_{l_1})}(k))\geq w_{e(b_{l_1})}$ 
and $|S_{v(b_{l_1})}(k)|\geq |T(b_{l_1},b_{l_2},\ldots, T_{b_{l_m}})|+\Path(v(b_{l_1})$ 
since the strategy ends at $v_t\not\in T(b_{l_1},b_{l_2},\ldots, T_{b_{l_m}})$. 
In the next step $S$ will move back to $\pred(v(b_{l_1}))$ along $e(b_{l_1})$ 
and in $S$ the root $v(b_{l_1})$ of the tree 
$T(b_{l_1},b_{l_2},\ldots, T_{b_{l_m}})$ and the edge $e(b_{l_1})$  will never be visited again. 

If we consider a strategy $S'$ that  first visits $v(b_{l_1})$, fully 
explores the subtree $T(b_{l_1},b_{l_2},\ldots, T_{b_{l_m}})$ by DFS and 
moves back to the start $v_s$ by passing $e(b_{l_1})$, the minimal number of agents required for this movement is exactly
$|T(b_{l_1},b_{l_2},\ldots, T_{b_{l_m}})|+\Path(v(b_{l_1})+w_{e(b_{l_1})}$ with  $w_{e(b_{l_1})}$
non-settled agents. With $$\left(|S_{v(b_{l_1})}|-|T(b_{l_1},b_{l_2},\ldots, T_{b_{l_m}})|-\Path(v(b_{l_1})\right)+w_{e(b_{l_1})}$$ 
agents we now start the whole sequence $S$ again. 

In the concatenation of $S'$ and $S$, say $S'S$,  the vertex 
$v(b_{l_1})$ is visited $k'=k+2$ times for $|T(b_{l_1},b_{l_2},\ldots, T_{b_{l_m}})|\geq 2$ and 
$k'=k+1$ times for $m=1$ and $v(b_{l_1})=T(b_{l_1},b_{l_2},\ldots, T_{b_{l_m}})$. 

After $S'$ was executed for the remaining movement of $S'S_{v(b_{l_1})}(k')$
the portion $w_{e(b_{l_1})}$ of $\left(|S_{v(b_{l_1})}|-|T(b_{l_1},b_{l_2},\ldots, T_{b_{l_m}})|-\Path(v(b_{l_1})\right)+w_{e(b_{l_1})}$ allows us to cross all edges 
in $S'S_{v(b_{l_1})}(k')$ for \emph{free}, because $w_{e(b_{l_1})}$ is the maximal 
weight in the tree. 

Thus obviously $\curr(S'S_{v(b_{l_1})}(k'))=
\curr(S_{v(b_{l_1})}(k))$ holds and $S'S$ and $S$ require the same 
number of agents. 
In $S'S$ all visits of $T(b_{l_1},b_{l_2},\ldots, T_{b_{l_m}})$ made by $S$ 
were useless because the tree was already completely filled by 
$S'$. Skipping all these visits in $S$, we obtain a sequence $S''$ 
and finally $S'S''$ has the desired structure.~\qed

\paragraph{Proof (Theorem~\ref{TreeBack-theo})}
The strategy of the single group 
has to return back to the start vertex $v_s$ 
at the end. Therefore no subtree  
$T(b_{l_1},b_{l_2},\ldots, b_{l_m})$ contains the vertex $v_s$ 
visited last. Let us assume that $N_1$ is the optimal 
number of agents required for $T$. 

After the first 
application of Lemma~\ref{orderbyedge-lem} to the subtree 
$T(b_{l_1},b_{l_2},\ldots, b_{l_m})$ with greatest incoming edge
 weight $w_{e(b_{l_1})}$ we can move with at least $w_{e(b_{l_1})}$ agents 
 back to the root $v_s$ without loss by the strategy $S'$. 
 Let us assume that $N'_1$ agents return to the start.  
 
We simple set all 
node weights along the path from $v_s$ to $v(b_{l_1})$ to zero, cut off
the fully explored subtree $T(b_{l_1},b_{l_2},\ldots, b_{l_m})$ and obtain a 
tree $T'$. 
Note that the collected subtrees were  disjoint and apart from $T(b_{l_1},b_{l_2},\ldots, b_{l_m})$ the remaining collected subtrees will be the same in $T'$ and $T$.
By induction on the number of the subtrees 
in the \emph{remaining} problem $T'$ we can visit the collected subtrees in 
the order of the dominating edge weights. 

Note that the number of agents required for $T'$ might be less than $N_1'$ 
because the weight $w_{e(b_{l_1})}$ was responsible for $N_1'$. This makes
no difference in the argumentation. 

We consider the running time. By a simple DFS walk of $T$, we compute the disjoint
trees $T(b_{l_1},b_{l_2},\ldots, b_{l_m})$ implicitly by pointers 
to the root vertices $v(b_{l_1})$. For any vertex $v$, 
there is a pointer to its unique subtree $T(b_{l_1},b_{l_2},\ldots, b_{l_m})$ and 
we compute the sum of the vertex weights for any subtree. 
This can be done in overall linear time. 
Finally, we can sort the trees by the order of the weights of the 
incoming edges in $O(n \log n)$ time for real weights and in 
$O(n)$ time for integer weights.

For computing the number of agents required, we make use of the following efficient 
procedure, similar to the algorithm indicated at the 
beginning of this Section. 
Any visited vertex will be marked. In the beginning let $\add:=0$ and $\curr:=N$. 
Let $|T(b_{l_1},b_{l_2},\ldots, b_{l_m})|$ 
denote the sum of the vertex weights of the corresponding tree.
We successively \emph{jump} to the vertices $v(b_{l_1})$ of the trees $T(b_{l_1},b_{l_2},\ldots, b_{l_m})$ 
by making use of the pointers. We mark $v(b_{l_1})$ and 
starting with the predecessor of $v(b_{l_1})$ we move backwards along the path from $v(b_{l_1})$ to the root $v_s$,  
until the first marked vertex is found.
Unmarked vertices along this path are labeled as marked 
and the sum of the corresponding vertex weights is 
counted in a variable $\Path$. Additionally, for any such vertex $v$ that belongs to 
some other subtree $T(\ldots)$ we subtract the 
vertex weight $w_v$ from $|T(\ldots)|$, this part of $T(\ldots)$ is already visited. 
 
Now we set $\curr:=\curr-(|T(b_{l_1},b_{l_2},\ldots, b_{l_m})|+\Path)$. 
If $\curr<w_e$ holds, we set $\add:=\add+(w_e-\curr)$ and 
$\curr:=w_e$ as before. Then we turn over to the next tree. Obviously 
with this procedure we compute the optimal number of agents in linear time, any vertex is marked only once. 
A pseudocode is presented in  Algorithm~\ref{eval2-alg}.\qed

We present an example of the execution of Algorithm~\ref{eval2-alg}.
For example in  Figure~\ref{exampleII-fig} we have $N:=41$ and first jump to the root 
$v_4$ of $T(b_6,b_7)$, we have $|T(b_6,b_7)|=8$. Then we count the $6$ agents along the 
path from $v_4$ back to $v_s$ and mark the vertices $v_2$ and $v_s$ 
as visited. This gives $\curr:=41-(8+6)=27$, which is greater than  $w_{e_7}=12$. 
Additionally, for $v_2$ we subtract $2$ from $|T(b_5)|$ which gives $|T(b_5)|=9$. 
Now we jump to the root $v_3$ of $T(b_2,b_3,b_4)$ with 
$|T(b_2,b_3,b_4)|=8$. Moving from $v_3$ back to $v_s$ to the first unmarked 
vertex just gives no step. No agents are counted along this path. 
Therefore $\curr:=27-(8+0)=19$ and $\curr>w_{e_5}=10$. 
Next we jump to the root $b_1$ of $T(b_1)$ of size $|T(b_1)|=3$. 
Moving back to the root we count the weight $5$ of the unvisited vertex $v_1$ (which will be marked now). Note that $v_1$ does not belong to a subtree $T(\ldots)$.
We have $\curr:=19-(3+5)=11$. Now we jump to the root $v_2$ of $T(b_5)$ of current size 
$|T(b_5)|=9$. Therefore $\curr:=11-(9+0)=2$ which is 
now smaller than $w_{e_4}=7$. This gives $\add:=\add+(w_e-\curr)=0+(7-2)=5$ and $\curr:=w_e=7$.
Finally we jump to $b_0=T(b_0)$ and have  $\curr:=7-(2-0)=5$ which is greater than $w_{e_2}$. 
Altogether $5$ additional agent can move back to $v_s$ and $N+\add=46$ agents are required in 
total. 

\begin{algorithm}
\caption{Return variant. Number of agents for $T=(V,E)$. 
  Roots $v(b_{l_1})$ of trees $T(b_{l_1},b_{l_2},\ldots, b_{l_m})$ are given by pointers in a list $L$ in the order of dominating edge weights. 
 $\NIL$ is the predecessor of root $v_s$.}
\label{eval2-alg}
\begin{algorithmic}
\STATE $N:=\sum_{v\in V} w_v$; $\curr:=N$; $\add:=0$; 
\WHILE {$L \neq \emptyset$}
\STATE $v(b_{l_1}):=\first(L)$; $\deleteFirst(L)$;
\STATE Mark $v(b_{l_1})$;
\STATE $\Path:=0$; $\currvert:=\pred(v(b_{l_1}))$;
\WHILE {$\currvert$ not marked and $\currvert\neq \NIL$}
        \STATE $\Path:=\Path+w_{\currvert}$;
        \IF {$\currvert$ belongs to $T(\ldots)$} 
                \STATE $|T(\ldots)|:=|T(\ldots)|-w_{\currvert}$
        \ENDIF
        \STATE Mark $\currvert$; $\currvert:=\pred(\currvert)$;
\ENDWHILE
\STATE $\curr:=\curr-(|T(b_{l_1},b_{l_2},\ldots, b_{l_m})|+\Path)$. 
\IF {$\curr< w_e$} \STATE $\add:=\add+(w_e-\curr)$; $\curr:=w_e$; 
\ENDIF
\ENDWHILE
\STATE \mbox{\bf\small RETURN} $N+\add$
\end{algorithmic}
\end{algorithm}

\subsection{Lower bound for traversal steps}\label{LBTraverse-sec}

\begin{figure}[hbtp]%
  \begin{center}%
    \includegraphics[scale=0.55,keepaspectratio]{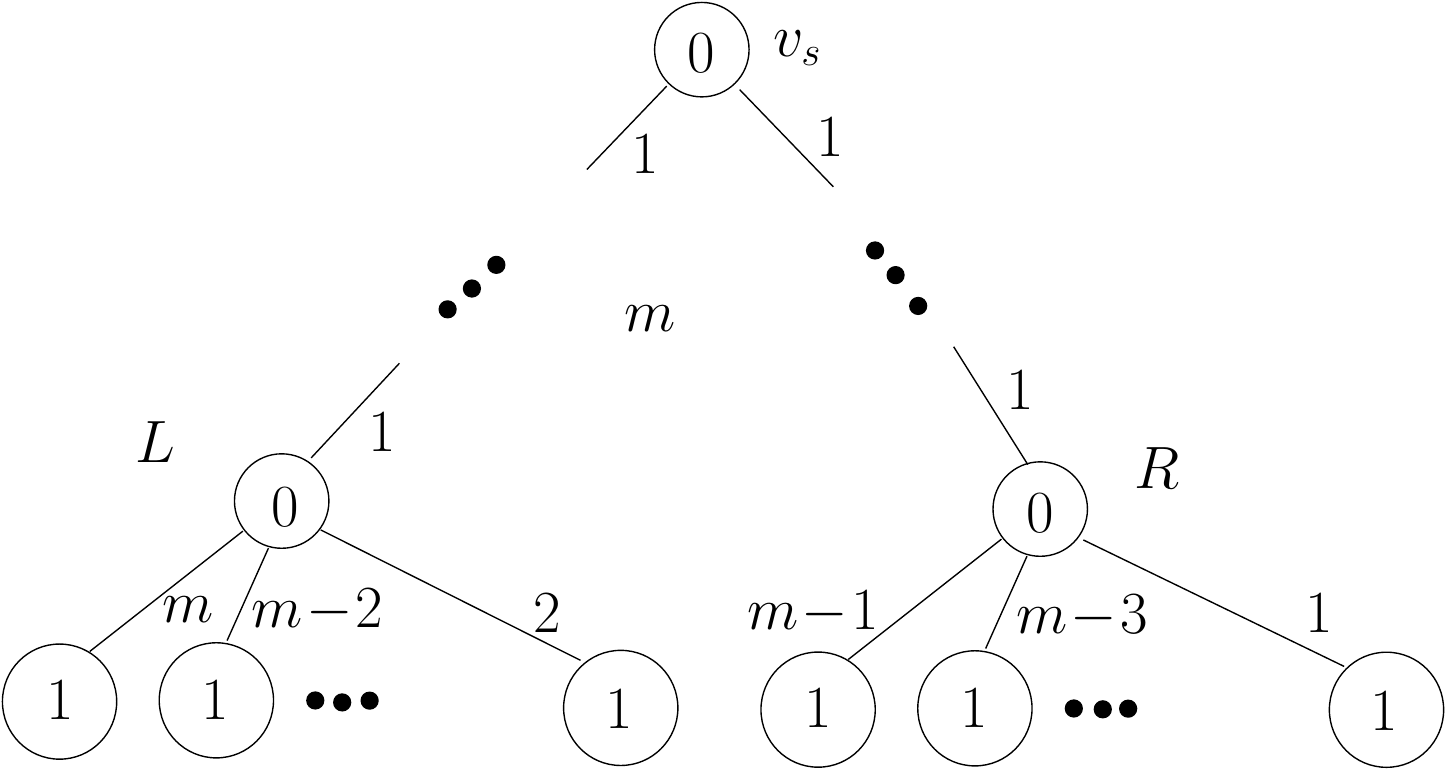}%
    \caption{An optimal deployment strategy for the tree with $3m$ edges 
requires $m+1$ agents and successively moves from $L$ to $R$ beyond 
$v_s$ in total $\Omega\left( m \over 2 \right)$ times. Thus $\Omega(m^2)$ steps are
required. }
    \label{worstcase-fig}
  \end{center}%
\end{figure}

It is easy to see that although the number of agents required and the visiting 
order of the leafs can be computed  sub-quadratic optimal time, the number of traversal steps for a tree could be in $\Omega(n^2)$; see the example in Figure~\ref{worstcase-fig}.  In this example the strategy with the minimal number of agents 
is unique and  the agents have to run in a single group. 

\subsection{No-return variant}\label{OptTreeNoReturn-sec}                  

Finally, we discuss the more difficult task of the no return variant. 
In this case for an optimal solution not all collected subtrees will be visited in 
the order of the decreasing dominating edge weights. 

For example a strategy for the no-return in Figure~\ref{exampleII-fig}
that visits the collected subtrees $T(b_6,b_7)$, $T(b_2,b_3,b_4)$, $T(b_1)$, $T(b_5)$ and  
$T(b_0)$ in the order of the weights $w_{e_7}=12$, $w_{e_5}=10$, $w_{e_3}=9$,
$w_{e_4}=7$ and $w_{e_2}=4$ of the dominating edges requires 
$46$ agents even if we do not finally move back to the start vertex. 
As shown at the end of Section~\ref{Return-sec} we required 
$5$ additional agents for leaving $T(b_5)$, entering and leaving $T(b_0)$ afterwards 
requires no more additional agents. 

In the no return variant, we can assume that any strategy ends in a leaf, because the last vertex that will be served has to be a leaf. This also means that it is reasonable 
to enter a collected subtree, which will not be left any more. 
In the example above we simply change the order of the last two
subtrees. If we enter the collected subtrees in the order $T(b_6,b_7)$, $T(b_2,b_3,b_4)$, $T(b_1)$, $T(b_0)$ and  
$T(b_5)$ and $T(b_5)$ is not left at the end, we end the strategy in $b_5$ (no-return) and 
require exactly $N=41$ agents, which is optimal.

\begin{theorem}\label{TreeGeneral-theo} 
For a weighted tree $T$ with given root 
$v_s$ and non-fixed end vertex we 
can compute an optimal visiting order of the leafs and 
the number of agents required in amortized time 
$O(n \log n)$.
\end{theorem}
For the proof of the above statement we first characterize the 
structure of an optimal strategy. Obviously we can assume that a strategy 
that need not return to the start will end in a leaf. 
Let us first assume that the final leaf, $b_t$, is already known or given. 
As indicated for the example above, the \emph{final} collected subtree will break the 
order of the collected subtrees in an optimal solution. 
This behaviour holds recursively. 

\setcounter{theorem}{7}
\begin{lemma}\label{lasttree-lem}  
An optimal traversal strategy that has to visit the leaf $b_t$ last can be 
computed as follows: Let $T(b_{l_1},b_{l_2},\ldots,b_{l_m})$ 
be the collected subtree of $T$ that contains~$b_t$. 
\begin{enumerate}
\item First, all collected subtrees 
$T(b_{q_1},b_{q_2},\ldots,b_{q_o})$ of the tree $T$ with dominating 
edge weight greater than  $T(b_{l_1},b_{l_2},\ldots,b_{l_m})$
are successively visited and fully explored (each by DFS) and left  in the decreasing order of the weights of the dominating edges.  
\item Then, the remaining collected subtrees that do not contain $b_t$
are visited in an arbitrary order (for example by DFS). 
\item Finally, 
the collected subtree $T(b_{l_1},b_{l_2},\ldots,b_{l_m})$ that contains $b_t$
is visited. Here we recursively apply the same strategy to
the subtree $T(b_{l_1},b_{l_2},\ldots,b_{l_m})$. That is, we 
build a list of collected subtrees for the tree $T(b_{l_1},b_{l_2},\ldots,b_{l_m})$ 
and recursively visit the collected subtrees by steps 1. and 2. so that the 
collected subtree that contains $b_t$ is recursively visited last in step 3. again.  
\end{enumerate}
\end{lemma}
\paragraph{Proof:} The precondition of the Theorem says that there is an optimal 
strategy given by a sequence $S$  of visited vertices and edges so that 
the strategy ends in the leaf $b_t$. 
Let $T(b_{l_1},b_{l_2},\ldots,b_{l_m})$ 
be the collected subtree of $T$ that contains~$b_t$ and 
let $w_{e(b_t)}$ be the corresponding dominating edge weight. 
So $b_t\in \{ b_{l_1},b_{l_2},\ldots,b_{l_m}\}$ and 
$v(b_t)$ is the root of  $T(b_{l_1},b_{l_2},\ldots,b_{l_m})$. 
Similarily as in the proof of Lemma~\ref{orderbyedge-lem} we would like to reorganize $S$ 
as required in the Lemma. 

For the trees $T(b_{q_1},b_{q_2},\ldots, b_{q_o})$ 
with dominating edge weight greater than $w_{e(b_t)}$ we can 
successively apply Lemma~\ref{orderbyedge-lem}. So we reorganize 
$S$ is this way by a sequence $S'$ that finally moves the agents back to the start vertex $v_s$. Then we apply the sequence $S$ again but skip the visits of 
all collected subtrees already fully visited by $S'$ before. 
This show step 1. of the Theorem. 

This gives an overall sequence $S'S''$ with the 
same number of agents and $S''$ does only visit collected subtrees of $T$ with 
dominating edges weight smaller than or equal to $w_{e(b_t)}$. 
Furthermore, $S''$ also ends in $b_t$. 

The collected subtree $T(b_{l_1},b_{l_2},\ldots,b_{l_m})$ with weight $w_{e(b_t)}$ 
does not contain any collected subtree with weight smaller than or equal to 
$w_{e(b_t)}$. At some point in $S''$ the vertex $v(b_t)$ is visited for the last time, say for the $k$-th time, 
by a movement from the predecessor $\pred(v(b_t)$ of  $v(b_t)$ by passing 
the edge of weight $w_{e(b_t)}$. At least $w_{e(b_t)}$ agents are still required for 
this step. At this moment all subtrees different from $T(b_{l_1},b_{l_2},\ldots,b_{l_m})$ 
and edge weight smaller than or equal to $w_{e(b_t)}$ habe been visited 
since the strategy ends in $b_t\in \{  b_{l_1},b_{l_2},\ldots,b_{l_m}\}$. 

Since $w_{e(b_t)}$  agents are required for the final movement along $e(b_t)$
there will be no loss of agents, if we postone all movements 
into $T(b_{l_1},b_{l_2},\ldots,b_{l_m})$ in $S''$ first and then 
finally solve the problem in $T(b_{l_1},b_{l_2},\ldots,b_{l_m})$ optimally. 
For the subtrees different from $T(b_{l_1},b_{l_2},\ldots,b_{l_m})$ 
and edge weight smaller than or equal to $w_{e(b_t)}$ we only require the 
agents that have to be placed there, since at least $w_{e(b_t)}$ non-settled 
agents will be always present.  
Therefore we can also decide to visit the subtrees different from $T(b_{l_1},b_{l_2},\ldots,b_{l_m})$ 
and edge weight smaller than or equal to $w_{e(b_t)}$ in an arbitrary order 
(for example by DFS).
This gives step 2. of the Theorem.

Finally, we arrive at $v(b_t)$ and $T(b_{l_1},b_{l_2},\ldots,b_{l_m})$ and would like to  end 
in the leaf $b_t$. By induction on the height of the trees the tree
$T(b_{l_1},b_{l_2},\ldots,b_{l_m})$ can be handled in the same way. 
That is, we build a list of collected subtrees for the tree $T(b_{l_1},b_{l_2},\ldots,b_{l_m})$ 
itself and recursively visit the collected subtrees by steps 1. and 2. so that the 
collected subtree that contains $b_t$ is recursively visited last in step 3. again. \qed

The remaining task for the proof of Theorem~\ref{TreeGeneral-theo}  is to efficiently find the
best leaf $b_t$ where the overall optimal strategy 
ends. The above Lemma states that we 
should be able to start the algorithm recursively at the root of 
a collected subtree $T(b_{l_1},b_{l_2},\ldots,b_{l_m})$ 
that contains $b_t$. For $v_s$ a list, $L$, of the collected subtrees for $T$ is given and 
for finding an optimal strategy we 
have to compute the corresponding lists of collected subtrees for all trees 
$T(b_{l_1},b_{l_2},\ldots, b_{l_m})$ in $L$ recursively.  

\begin{figure}[hbtp]%
  \begin{center}%
    \includegraphics[scale=0.53,keepaspectratio]{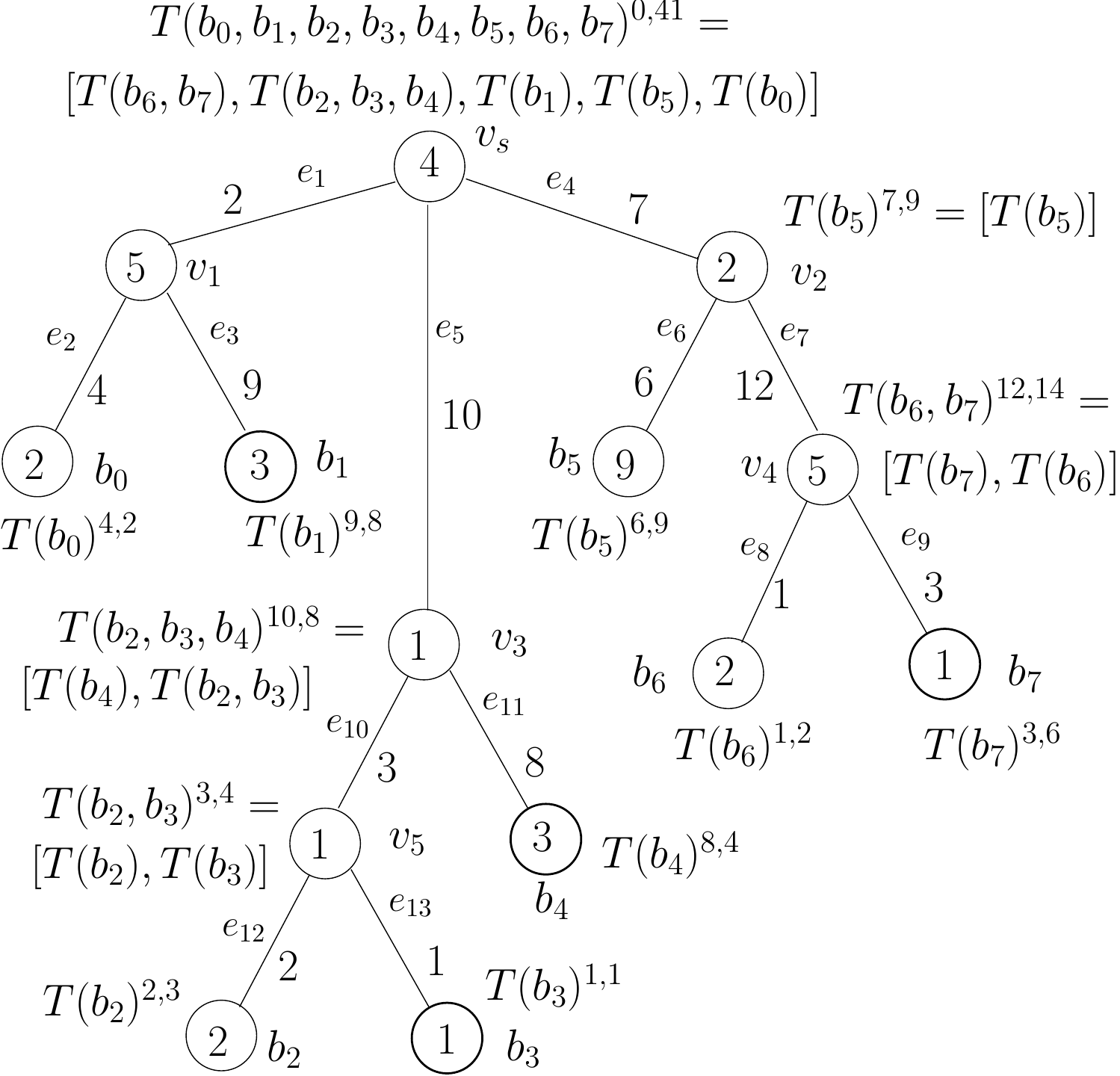}%
    \caption{All information required can be computed recursively 
from bottom to top in amortized $O(n \log n)$ time.}
    \label{exampleIII-fig}
  \end{center}%
\end{figure}

Figure~\ref{exampleIII-fig} shows an example. 
In this setting let us for example consider the case that we would like to 
compute an optimal  visiting order so that the strategy has to end in the leaf $b_2$. 
Since $b_2$ is in $T(b_2,b_3,b_4)$ in the list of $v_s$ in Figure~\ref{exampleIII-fig} by the above Lemma  in step 1. we first visit the tree $T(b_6,b_7)$ of dominating edge weight greater than the dominating edge weight of $T(b_2,b_3,b_4)$. Then we visit 
$T(b_1)$, $T(b_5)$ and $T(b_0)$ in step 2. After that in step 3.  we recursively start the 
algorithm in $T(b_2,b_3,b_4)$. Here at $v_3$ the list of collected sutrees contains 
$T(b_4)$ and $T(b_2,b_3)$ and by the above recursive algorithm in step 1. we 
first visit $T(b_4)$. There is no tree for step 2. and we recursively enter $T(b_2,b_3)$ 
at $v_5$ in step 3. Here for step 1. there is no subtree and we enter the 
tree $T(b_3)$ in step 2. until finally we recursively end in $T(b_2)$ in step 3. 
Here the algorithm ends. Note that in this example $b_2$ is not the overall optimal 
final leaf. 

If we simply apply the given algorithm for any leaf and compare the given 
results (number of agents required) we require $O(n^2 \log n)$ computational time. 
For efficiency, we compute the required information in a single step and check the 
value for the different leafs successively. 
It can be shown that in such a way the best leaf $b_t$ and the overall optimal strategy can be computed in amortized  
$O(n \log n)$ time.

%

Finally, we give the proof for Theorem~\ref{TreeGeneral-theo}  by the following discussion. 
We would like to compute the lists of the collected subtrees 
$T(b_{l_1},b_{l_2},\ldots,b_{l_m})$ recursively. 
More precisely, for the root $v_s$ of a full tree $T$ with leafs $\{b_1,b_2,\ldots,b_n\}$
we obtain a list,  denoted by $T(b_1,b_2,\ldots,b_n)$,  of the 
collected subtrees of $T$ with respect to the decreasing order of the 
dominating edge weights as introduced in Section~\ref{Algorithm-sec}. 

The elements of the list are pointers to the roots of the collected subtrees $T(b_{l_1},b_{l_2},\ldots, b_{l_m})$. 
For any such root $T(b_{l_1},b_{l_2},\ldots, b_{l_m})$ of a subtree in the list $T(b_1,b_2,\ldots,b_r)$ 
we recursively would like to compute the corresponding list of collected subtrees  recursively; see Figure~\ref{exampleIII-fig} for an example. 

Additionally, for any considered collected subtree  $T(b_{k_1},b_{k_2},\ldots, b_{k_r})$ 
that belongs to the pointer list of $T(b_{l_1},b_{l_2},\ldots, b_{l_m})$ 
we store a pair of integers $x,y$ at the corresponding root 
of $T(b_{k_1},b_{k_2},\ldots, b_{k_r})$ ; see Figure~\ref{exampleIII-fig}.
Here $x$ denotes the weight of the 
dominating edge. The value $y$ denotes the size of $|T(b_{k_1},b_{k_2},\ldots, b_{k_r})|+\Path)$,  
if we recursively start the optimal tree algorithm
in the root $T(b_{l_1},b_{l_2},\ldots, b_{l_m})$; see Algorithm~\ref{eval2-alg}.  
This means that $y$ denotes the size of the collected subtree \emph{and} 
the sum of the weights along the path back from $T(b_{k_1},b_{k_2},\ldots, b_{k_r})$ 
to the root $T(b_{l_1},b_{l_2},\ldots, b_{l_m})$ if $T(b_{k_1},b_{k_2},\ldots, b_{k_r})$, 
if $T(b_{k_1},b_{k_2},\ldots, b_{k_r})$ is the first entry of  the list 
$T(b_{l_1},b_{l_2},\ldots, b_{l_m})$ and therefore has maximal weight. 

The list of subtrees at the root $v_s$  of $T$ is denoted by $T(b_{1},b_{2},\ldots, b_{r})^{x,y}$ 
and obtains the values $x:=0$ (no incoming edge) and $y:=N$ (the sum 
of the overall vertex weights). 
We can show that all information
can be computed efficiently from bottom to top and finally also allows us to compute an overall 
optimal strategy. 

For the overall construction of all pointer lists $T(b_{l_1},b_{l_2},\ldots, b_{l_m})$ we 
internally make use of Fibonacci heaps~\cite{ft-fhtui-87}. 
The corresponding heap for a vertex $v$ always contains \emph{all} collected subtrees 
of the leafs of $T_v$. The collected subtree list for the vertex $v$ itself might be empty; 
see for example that vertex $v_1$ does not root a set of collected subtree. 
In the following the list of pointers to collected subtrees is denoted by $[\ldots]$ and the 
internal heaps are denoted by $(\ldots)$. 

The subtrees in the heap are also given by pointers. But the heap is sorted by 
 \emph{increasing} dominating edge weights. Note, that we have two 
different structures. Occasionally a final subtree for a vertex with a list of pointers for the collected subtrees (in decreasing order)  
and the internal heaps with a collection of all collected subtrees (in increasing order) 
have the same elements.

With the help of the heaps we successively compute and store the final collected 
subtree lists for the vertices. We start the computations on the leafs of the tree. 
For a single leaf $b_l$ 
the heap $(T(b_l)^{x,y})$ and the subtree $T(b_l)^{x,y}$ represent exactly the same. The value
$x$ of  $T(b_l)^{x,y}$ is given by the edge weight of the leaf. The value $y$ of  $T(b_l)^{x,y}$
will be computed recursively, it is initialized by the 
vertex weight of the leaf. For example, in Figure~\ref{exampleIII-fig} for $b_7$ 
and $b_6$ we first have $T(b_6)^{1,2}$ and $T(b_7)^{3,1}$, representing both the heaps and the subtrees. 

Let us assume that the heaps for the child nodes of an  
internal node $v$ already have been computed and $v$ is a branching vertex with 
incoming edge weight $w_e$. We have to add the node weight of $v$  
to the value $y$ of one of the subtrees in the heap. We simply additionally store 
the subtree with greatest weight among the branches. 
Thus in constant tim 
 we add the node weight of the branching 
vertex to the value $y$ of a subtree with greatest weight. 
Then we unify the heaps of the children. 
They are given in the \emph{increasing} order of the dominating edges weights. 
This can be done in time proportional to the number of child nodes of $v$. 
 For example, at $v_4$ in Figure~\ref{exampleIII-fig} first we increase the $y$-element of 
the subtree $T(b_7)^{3,1}$ in the heap
by the vertex weight $5$ of $v_4$ which gives $T(b_7)^{3,6}$.  
Then we unify $(T(b_6)^{1,2})$ and $(T(b_7)^{3,6})$ to a heap 
$(T(b_6)^{1,2},T(b_7)^{3,6})$. For convenience in the heap we attach the values $x$ and $y$ directly to the pointer of the subtree. 

Now, for branching vertex $v$ by using the new unified heap we find, delete and collect the subtrees with 
minimal incoming edge weight as long as the weights are 
smaller than or equal to the weight $w_e$. If there is no such tree, 
we do not have to build a new collected subtree at this vertex and 
also the heap remains unchanged. 
If there are some subtrees that have incoming edge weight smaller than or equal to $w_e$
the pointers to all these subtrees will build a new collected 
subtree $T(b_{l_1},b_{l_2},\ldots,b_{l_m})$ with $x$-value $w_e$ at the node $v$. 
Additionally, the pointers to the corresponding subtrees of $T(b_{l_1},b_{l_2},\ldots,b_{l_m})$
can easily be ordered with \emph{increasing} weights since we have 
deleted them out of the heap starting with the smallest weights.  
Additionally, we sum up the values $y$ of the deleted subtrees. 
Finally, we have computed the collected 
subtree $T(b_{l_1},b_{l_2},\ldots,b_{l_m})$ and 
its information $x,y$ at node $v$. At the end a new subtree is also inserted into the fibonacci heap of the 
vertex $v$ for future unions and computations.

For example in Figure~\ref{exampleIII-fig} for the just computed heap $(T(b_6)^{1,2},T(b_7)^{3,6})$ at vertex $v_4$ 
we delete and collect the subtrees $T(b_6)^{1,2}$ and $T(b_7)^{3,6}$ out of the 
heap because the weight $w_{e_7}=12$ dominates both weights $1$ and $3$. 
This gives a new subtree $T(b_7,b_6)^{12,8}=[T(b_7),T(b_6)]$ at $v_4$ and also a heap 
$(T(b_7,b_6)^{12,8})$. 

Note, that sometimes no new subtree is build if no tree is deleted out of the 
heap because the weight of the incoming edge is less than the current weights. Or it might happen that 
only a single tree of the heap is collected and gets a new dominating edge. 
In this case also no subtree is deleted out of the heap. 
We have a single subtree with the same leafs as before but with a different 
dominating edge. We do not build a 
a collected subtree for the vertex at this moment, the insertion of such subtrees at the corresponding vertex is postponed. 
 
For example for the vertex $v_2$ with incoming edge weight $7$ 
in Figure~\ref{exampleIII-fig} we have already 
computed the heaps $(T(b_5)^{6,9})$ and $(T(b_6,b_7)^{12,8})$ of the subtrees. 
Now the vertex weight $2$ of $v_2$ is added to the $y$-value of $T(b_6,b_7)^{12,8}$ 
which gives $T(b_6,b_7)^{12,10}$ for this subtree.
Then we unify the heaps to $(T(b_5)^{6,9},T(b_6,b_7)^{12,10})$. 
Now with respect to the incoming edges weight $7$ only the first tree in the 
heap is collected to a subtree and this subtree gives the list for vertex $v_2$. 
The heap of the vertex $v_2$ now reads $(T(b_5)^{7,9},T(b_7,b_6)^{12,10})$ 
and the collected subtree is $T(b_5)^{7,9}=[T(b_5)]$.

Finally, we arrive the root vertex $v_s$ and all subtrees of the heap 
are inserted into the list of collected subtrees for the root. 


The delete operation for the heaps requires amortized $O(\log m)$ time for 
a heap of size $m$ and subsumes any other operation. 
Any delete operation leads to a collection 
of subtrees, therefore at most $O(n)$ delete operation will 
occur. Altogether all subtrees and its pointer lists and the values $x$ and $y$ 
can be computed in amortized $O(n \log n)$ time.

The remaining task is that we use the information of the subtrees for 
calculating the optimal visiting order of the leafs in overall 
$O(n \log n)$ time. Here Algorithm~\ref{eval2-alg} will be used as a subroutine. 

As already mentioned we only have to fix the leaf $b_t$ 
visited last. We proceed as follows. 
An optimal strategy ends in a given 
collected subtree with some dominating edge weight $w_e$. 
The strategy visits and explores the remaining trees 
in the order of the dominating edges weights. 

Let us assume that on the top level the 
collected subtrees are ordered by the weights $w_{e_1}\leq w_{e_2}\leq\ldots\leq w_{e_j}$.
Therefore by the given information and with Algorithm~\ref{eval2-alg} 
for any $i$ we can successively 
compute the number of additional agents required for any successive order 
$w_{e_{i+1}}\leq w_{e_{i+2}}\leq\ldots\leq w_{e_j}$
and by the 
$y$-values we can also 
compute the number of agents required for the trees of the 
weights $w_{e_1}\leq w_{e_2}\leq\ldots\leq w_{e_{i-1}}$. 
The number of agents required for the final tree of weight $w_{e_{i}}$ and the 
best final leaf stems from recursion. 
With this informations the number of agents can be computed. 
This can be done in overall linear time $O(j)$ for any $i$. 

The overall number of collected subtrees in the construction is linear 
for the following reason.
We start with $n$ subtrees at the leafs. If this subtree appears again in some 
list (not in the heap), either it has been collected together with some others or it 
builds a subtree for its own (changing dominance of a single tree). 
If it was collected, it will never appear for its 
own again on the path to the root. If it is a single subtree of that node, 
no other subtree appears in  the list at this node. Thus for the $O(n)$ nodes we have 
$O(n)$ collected subtrees in the lists total. 

From $i$ to $i+1$ only a constant number of additional calculations have 
to be made. By induction this can recursively be done for the subtree 
dominated by $w_{e_i}$ as well. Therefore we can use the 
given information for computing the optimal strategy 
in overall linear time $O(n)$ if the collected subtrees are given recursively.

\section{Conclusion}\label{conclu-sec}                

We introduce a novel traversal problem in weighted graphs that 
models security or occupation constraints and gives rise to many further 
extensions and modifications. 
The problem discussed here is NP-hard in general and can be solved efficiently for 
trees in $\Theta(n\log n)$ where some machinery is necessary. This also gives a $2$-approximation for 
a general graph by the MST.


\bibliographystyle{plain}

\end{document}